\def\be{\begin{equation}}
\def\te{\end{equation}}
\def\ba{\begin{eqnarray}}
\def\nn{\nonumber}
\def\ta{\end{eqnarray}}

\def\a{\alpha}
\def\b{\beta}
\def\c{\raisebox{.4ex}{$\chi$}}
\def\d{\delta}
\def\e{\epsilon}

\def\k{\kappa}
\def\l{\lambda}
\def\m{\mu}
\def\n{\nu}

\def\p{\pi}

\def\r{\rho}
\def\s{\sigma}

\def\O{\Omega}

\newskip\humongous \humongous=0pt plus 1000pt minus 1000pt

\newif\ifdtup

\def\a{\alpha}

\def\ha{{1\over 2}}

\def\(#1){(\ref{#1})}

\documentstyle[12pt]{article}

\makeatletter                    
\@addtoreset{equation}{section}  
\makeatother                     

\begin{document}

\title{Notes on Black Hole Fluctuations and Backreaction
\thanks{Published in {\it Black Holes, Gravitational Radiation and
the Universe: Essays in honor of  C. V. Vishveshwara},
eds. B. R. Iyer and B. Bhawal  
(Kluwer Academic Publishers, Dordrecht, 1999) pp. 103-120}}
\author{B. L. Hu\\
{\small Department of Physics, University of Maryland,
 College Park, Maryland 20742, USA}\\
Alpan Raval\\
{\small Department of Physics, University of Wisconsin-Milwaukee,
Milwaukee, Wisconsin 53201, USA }\\
Sukanya Sinha\\
{\small Raman Research Institute, Bangalore, India}}
\date{\small {\it (umdpp 98-81, Feb 2, 1998)}}
\maketitle

\begin{abstract}
In these notes we prepare the ground for a systematic investigation into 
the issues of black hole fluctuations and backreaction by discussing the 
formulation of the problem, commenting on possible advantages and
shortcomings of existing works,  and introducing our own approach via 
a stochastic semiclassical theory of gravity based on  the 
Einstein-Langevin equation and  the fluctuation-dissipation relation 
for a self-consistent description of metric  fluctuations and dissipative 
dynamics of the black hole with backreaction from its Hawking radiance.
\end{abstract}

\newpage

\section{Classical and Semiclassical Backreaction of Metric Perturbations}

The idea of viewing a black hole (particle detector) interacting with a 
quantum field as a dissipative system, and the Hawking  \cite{Haw}- Unruh
\cite{Unr}  radiation as a manifestation of a  fluctuation-dissipation
relation was first proposed by Candelas and Sciama
\cite{CanSci,Sciama}. Even though, as we will soon see,  the fluctuations 
in the thoughts of these earlier authors are not the correct ones and the
relations proposed not really addressing
the backreaction of quantum fields in a classical black hole spacetime, the 
idea remains attractive. Indeed one of us (BLH) found it so attractive
that he launched a systematic 
investigation into the statistical mechanical properties of particle/spacetime 
and  quantum field  interactions. This involved the introduction of
statistical mechanical ideas
such as quantum open sytems \cite{qos} and field-theoretical methods such as the 
influence functional \cite{if} and Schwinger-Keldysh formalisms \cite{ctp} 
for the  establishment of a quantum statistical field theory in curved
spacetime (for a review, 
see \cite{HuPhysica,Banff}). It was found that the backreaction of quantum 
fields  (through processes like particle creation) on a classical
background spacetime can be 
described by an
Einstein-Langevin equation \cite{CH94,HM3}, which is a generalization of the 
semiclassical  Einstein equation  to include stochastic sources due to
created particles.
It was also found  from first principles that the backreaction can be
encapsulated in the form
of a Fluctuation-Dissipation Relation (FDR) \cite{HuSin,RHA}, which takes
into account
the mutual influence  of the quantum field and the background spacetime
(or detector,
in the case of Unruh radiation).
We expect it to hold  also for black hole systems, both in the familiar
static condition
where black hole thermodynamics based on the Bekenstein-Hawking relation
was constructed, and for dynamical collapse problems. This is the major theme
in our current program of research.

Note that it is nontrivial that such a relation exists at least on two counts. 
First, that the dynamics  of spacetime interacting with a quantum field
can indeed be treated 
like a classical particle  in a trajectory with stochastic components as
depicted in quantum 
Brownian  motion \cite{HPZ,GelHar2};  and second, although in statistical
thermodynamics 
the FDR is usually derived for  near-equilibrium conditions (via linear
response theory) , 
these semiclassical gravity processes can serve as an illustration that such a
relation can indeed exist for nonequilibrium processes.  This was
conjectured by one of
us in 1989 \cite{HuPhysica}
and demonstrated in succeeding work \cite{CH94,HM3,HuSin,CamVer}. We will
follow this line of thought to pursue  black hole backreaction problems.  
In this section 
we will describe in general terms the classical and semiclassical
backreaction of metric perturbations. In Sec. 2 we describe stochastic
semiclassical gravity in a cosmological setting, focusing on  the 
derivation of an Einstein-Langevin equation. In
Sec. 3 we describe metric fluctuations and backreaction in semiclassical
gravity, and the FDR for a black hole in equilibrium with its thermal  
radiation.
We comment on the inadequacy in Mottola's \cite{Mot} derivation of a FDR for
backreaction of static black holes. In Sec. 4 we discuss dynamical  black
holes,
identifying the Bardeen metric as a fruitful avenue for
semiclassical backreaction analysis. We also comment on the nature of
Candelas and Sciama's \cite{CanSci} FDR and its shortcoming for the
description of backreaction. We then summarize Bekenstein's theory
of black hole fluctuations and note the difference from our approach.
In these notes we are merely preparing the ground for our investigation
by sorting out the issues and identifying the inadequacies of previous
approaches. Details of our  findings will appear in later publications.

For brevity we shall use schematic expressions to discuss the
ideas here while relegating the details to research papers in progress.
Let us start with the classical Einstein equation for a spacetime with metric
$g_{\a\b}$
\be
G_{\mu \nu} (g_{\a \b}) = \k T_{\mu \nu}^{cm}
\te
where $G_{\m\n}$ is the Einstein tensor, $\k =8\p G_N$, $G_N$ being the Newton
constant, and  $T_{\mu \nu}^{cm}$ is the energy momentum tensor of some  
classical
matter (cm).
Consider perturbations of the metric tensor at the classical level, i.e.,

\be
g_{\a\b} = g_{\a\b}^{(0)} + \e h_{\a\b}^{(1)} + \e^2 h_{\a\b}^{(2)} +  ...
\te
where the superscript $n$ in parentheses on $h_{\a\b}^{(n)}$ indicates
the order of the perturbation. Most studies of gravitational waves
and instability are carried out for linear perturbations $n=1$.

The linear perturbations (in harmonic gauge) satisfy the linearized
 Einstein equations in the form \cite{MTW}

\be
\Box h_{\a\b} = 2 \k \d T_{\a\b}^{cm}
\te
The problem of gravitational perturbations in a cosmological
Friedmann-Lemaitre-Robertson-Walker (FLRW) spacetime in
relation to galaxy formation was first treated in detail by Lifshitz
\cite{Lifshitz}. Perturbations in a Schwarzschild black hole spacetime
was treated by Regge and Wheeler, Vishveshwara, Zerilli et al 
\cite{pertSchw}
and in a Kerr black hole by Teukolsky, Chandrasekhar and others 
\cite{pertKerr}.

For quantum fields in a classical background spacetime with metric
$g_{\m\n}^{(0)}$, the wave equation for, say, a massive ($m$) minimally  
coupled
scalar field $\Phi$ is

\be
\Box \Phi + m^2 \Phi =0
\te
The first order metric perturbations $h_{\a\b}^{(1)}$ in a vacuum
($T_{\a\b} = 0$)
obey an equation similar in form to the above with $m=0$ (the Lifshitz  
equation
\cite{Lifshitz}) and can thus be identified as two components (because   
of the 2
polarizations) of a massless minimally coupled scalar field.

Now let us consider the backreaction of gravitational perturbations or
quantum fields in classical and semiclassical gravity. At the classical
level, assuming a vacuum background, the linear perturbations
$h_{\mu \nu}^{(1)}$
would contribute a source to the $O(\e^0)$ equation in the form 
\cite{MTW}(Eq. 35.70)
\be
G_{\m\n} (g^{(0)}) = \k \langle T_{\m\n}^{gw}\rangle_I
\te
where
\be
T_{\m\n}^{gw} \equiv  \frac {1}{32\p}  [\bar h^{(1)}_{\a\b|\m}\bar
h^{(1)}_{\a\b|\n}],
\te
is the (inhomogeneous) energy momentum tensor of the gravitational waves (gw)
described (in a transverse-traceless gauge) by
$\bar h _{\a\b}^{(1)}\equiv h_{\a\b}^{(1)} - \ha h^{(1)} g_{\a\b}^{(0)}$.
Here the $\langle\, \rangle_I $ around $ T_{\m\n}^{gw} $
denotes the Isaacson average over the inhomogeneous sources
(taken over some intermediate wavelength
range larger than the natural wavelength of the waves but smaller than
the curvature radius of the background spacetime). This is an example of
backreaction at the classical level. (For related work see \cite{Tauber}.)

At the semiclassical level, the backreaction comes from particles
created in the quantum field (qf) on the background spacetime \cite{cpc,BirDav}.
Schematically the semiclassical Einstein equation takes the form
\be
\label{semi}
G_{\m\n} (g^{(0)}) = \k \langle T_{\m\n}^{qf}\rangle_V
\te
where
\be
T_{\m\n}^{qf} \equiv   \Phi,_\m \Phi,_\n + \ha m^2\Phi^2 g_{\m\n}
\te
is the energy momentum tensor of, say, a massive minimally coupled scalar
field
\footnote{By virtue of what we discussed above,  the gravitons --
quantized linear
perturbations of the background metric -- obey an equation similar in form
to that of a massless minimally coupled quantum scalar field. For graviton
production in cosmological spacetimes see \cite{Gri,ForPar}.}.
Here $\langle\,\rangle_V$ denotes expectation value taken with respect to
some vacuum state
with symmetry commensurate with that of the background spacetime.
Studies of semiclassical Einstein equation has been carried out in the last
two decades by many authors for cosmological
\cite{cpcbkr} and black hole spacetimes \cite{bhbkr}.

This is the point where our story begins. In the last decade we have been
able to move one step beyond in the semiclassical backreaction problem,
extending the above semiclassical framework to a stochastic semiclassical
theory, where noise and fluctuations from particle creation are accounted
for and incorporated. Spacetime dynamics is now governed by a stochastic
generalization of the semiclassical Einstein equation
known as the Einstein-Langevin equation --- the conventional theory of
semiclassical gravity with sources given by the vacuum expectation value
of the energy momentum tensor being a mean field approximation of this new
theory. We now describe the essence of this new theory, using gravitational
perturbations for illustration, again in a schematic form.  In this
context we can see the
distinction between metric perturbations and metric fluctuations on the one
hand and the proper meaning of metric fluctuations on the other.

\section{Stochastic Semiclassical Gravity: Einstein-Langevin Equation}

For concreteness let us consider the background spacetime to be a spatially
flat FLRW universe with metric $\tilde g_{\m\n}^{RW}$ plus small perturbations
$\tilde h_{\m\n}$,
\begin{equation}
   \tilde g_{\mu\nu}(x)=\tilde g_{\m\n}^{RW} + \tilde h_{\m\n}
                       \equiv e^{2\a(\eta)}g_{\mu\nu}
   \label{eq:FLRW}
\end{equation}
It is conformally related [with conformal factor  $\mbox{\rm
exp}(2\a(\eta))$] to
the Minkowski metric $\eta_{\mu\nu}$ and its perturbations $h_{\mu\nu}(x)$:
\be
g_{\m\n} =\eta_{\mu\nu}+h_{\mu\nu}(x)
\te
Here $\eta$ is the conformal time related to the cosmic time $t$ by
$dt=\mbox{\rm exp}[\a(\eta)]d\eta$. The perturbations $\tilde h_{\m\n}$
can be either anisotropic, as in a Bianchi type (Type I case is
treated by Hu and Sinha \cite{HuSin}), or inhomogeneous (treated
by Campos and Verdaguer \cite{CamVer}).  Here we follow these works.

The classical action for a free massless conformally coupled real
scalar field $\Phi(x)$ is given by
\be
S_f[\tilde g_{\mu\nu},\Phi]=
        -{1\over2}\int d^nx\sqrt{-\tilde g}
            \left[\tilde g^{\mu\nu}\partial_\mu\Phi\partial_\nu\Phi
                 +\xi(n)\tilde R\Phi^2
            \right],
\te
where $\xi(n)=(n-2)/[4(n-1)]$, and $\tilde R$ is the Ricci scalar
for the metric $\tilde g_{\mu\nu}$. Define a conformally related field
$\phi(x)\equiv \mbox{\rm exp}[(n/2-1)\a(\eta)]\Phi(x)$,
the action $S_f$ (after one integration by parts) 
\begin{equation}
   S_f[g_{\mu\nu},\phi] =
        -{1\over2}\int d^nx\sqrt{- g}
            \left[g^{\mu\nu}\partial_\mu\phi\partial_\nu\phi
                 +\xi(n) R\phi^2
            \right]
\end{equation}
takes the form of an action for a free massless conformally coupled real
scalar field $\phi(x)$ in a spacetime with metric $g_{\mu\nu}$, {\it i.e.}
a nearly flat spacetime. As the physical field $\Phi(x)$
is related to the field $\phi(x)$ by a power of the conformal factor
a positive frequency mode of the field $\phi(x)$ in flat spacetime will
correspond to a positive frequency mode in the conformally related
space. One can thus establish a quantum field theory in the conformally
related space by use of the conformal vacuum (see \cite{BirDav}).
Quantum effects such as particle creation and trace anomalies arise from
the breaking of conformal flatness of the spacetime
produced by the perturbations $h_{\mu\nu}(x)$.

The stochastic semiclassical Einstein equation differs from the semiclassical
Einstein equation (SCE) by a) the presence of a stochastic term measuring
the fluctuations of quantum sources
(arising from the difference of particles created in neighboring histories,
see, \cite{CH94}) and b) a dissipation term in the dynamics of spacetime 
(see the discussion following Eq. (2.7) for earlier treatments of metric 
dissipation).
Thus it endows the form of an Einstein-Langevin equation \cite{HM3}.
Two points are noteworthy: a) The fluctuations and dissipation (kernels)
obey a fluctuation -dissipation relation, which embodies the backreaction
effects of quantum fields on classical spacetime.
b) The stochastic source term engenders metric fluctuations.
The semiclassical Einstein equation depicts
a mean field theory which one can  retrieve from the Einstein-Langevin
 equation by
taking a statistical average with respect to the noise distribution.

The stochastic semiclassical Einstein equation, or Einstein-Langevin  
equation
takes on the form
\begin{eqnarray}
   \tilde G^{\mu\nu}(x)
        &=& \k  \left( T^{\mu\nu}_{c} +  T^{\mu\nu}_{qs} \right)
           \nonumber \\
   T^{\mu\nu}_{qs}
        &\equiv& \langle T^{\mu\nu} \rangle_{q} +  T^{\mu\nu}_{s}
           \nonumber \\
   T^{\mu\nu}_{s} &\equiv&  2e^{-6\alpha}F^{\mu\nu}[\xi]
   \label{eq:effective stress tensor}
\end{eqnarray}
Here, $T^{\mu\nu}_c$ is due to classical matter or fields,
$\langle T^{\mu\nu} \rangle_q$ is the vacuum expectation value of the
stress tensor of the quantum field, and  $T^{\mu\nu}_{qs}$ is a new  
stochastic
term. Up to first order in $h_{\mu\nu}$ they
are given by
\begin{eqnarray}
   \langle T^{\mu\nu}_{(0)} \rangle_q
        &=& \l\left[ \tilde H^{\mu\nu}_{(0)}
                        -{1\over 6} \tilde B^{\mu\nu}_{(0)}
                  \right]
           \nonumber \\
   \langle T^{\mu\nu}_{(1)} \rangle_q
        &=& \l\left[ \atop \right.
                         (\tilde H^{\mu\nu}_{(1)}
                          -2\tilde R^{(0)}_{\alpha\beta}
                          \tilde C^{\mu\alpha\nu\beta}_{(1)}
                         )
                        -{1\over 6} \tilde B^{\mu\nu}_{(1)}
              \nonumber \\
          & &\hskip1cm
                        +3e^{-6\alpha}
                            \left( -4( C^{\mu\alpha\nu\beta}_{(1)}
                                       \alpha
                                     )_{,\alpha\beta}
                        +\int d^4y A^{\mu\nu}_{(1)}(y)
                                      \mbox{\rm K}(x-y;\bar\mu)
                           \right)
                  \left. \atop \right].
   \label{eq:quantum stress tensor}
\end{eqnarray}
where the constant $\l = 1/2880\p^2$ characterizes one-loop quantum  
correction
terms (which include the trace anomaly and particle creation processes
\cite{Dew75}) and $\bar \mu$ is a renormalization parameter.
Here $C_{\m\a\n\b}$ is the Weyl curvature tensor and the
tensors $B^{\mu\nu}(x)$, $A^{\mu\nu}(x)$ and $H^{\mu\nu}(x)$ are
given by (see, e.g., \cite{CamVer,FulPar,Dew75} and earlier references)
\begin{eqnarray}
   B^{\mu\nu}(x)
      &\equiv& {1\over2}g^{\mu\nu}R^2
               -2RR^{\mu\nu}
               +2R^{;\mu\nu}
               -2g^{\mu\nu}\Box_g R,
         \nonumber \\
   A^{\mu\nu}(x)
      &\equiv& {1\over2}g^{\mu\nu}C_{\alpha\beta\rho\sigma}
                                  C^{\alpha\beta\rho\sigma}
               -2R^{\mu\alpha\beta\rho}{R^\nu}_{\alpha\beta\rho}
               +4R^{\mu\alpha}{R_\alpha}^\nu
         \nonumber \\
      & &\hskip .1cm
               -{2\over3}RR^{\mu\nu}
               -2\Box_g R^{\mu\nu}
               +{2\over3}R^{;\mu\nu}
               +{1\over3}g^{\mu\nu}\Box_g R,
         \nonumber \\
   H^{\mu\nu}(x)
      &\equiv& -R^{\mu\alpha}{R_\alpha}^\nu
               +{2\over3}RR^{\mu\nu}
               +{1\over2}g^{\mu\nu}R_{\alpha\beta}R^{\alpha\beta}
               -{1\over4}g^{\mu\nu}R^2.
\end{eqnarray}
All terms mentioned so far in the semiclassical Einstein equation,
including the dissipative kernel $K$, are familiar from backreaction  
calculations
done in the seventies and eighties (see. e.g., \cite{cpcbkr}). The new
result in the nineties is in the appearance of a stochastic source
\cite{HuSin,CamVer}, the tensor $F^{\mu\nu}(x)$
\begin{equation}
   F^{\mu\nu}(x)=
       -2\partial_\alpha\partial_\beta\xi^{\mu\alpha\nu\beta}(x),
   \label{eq:source}
\end{equation}
which is symmetric and traceless, {\it i.e.}
$F^{\mu\nu}(x)=F^{\nu\mu}(x)$ and $F^\mu_{\,\,\mu}(x)=0$ (implying
that there is no stochastic correction to the trace anomaly).
It accounts for the noise associated with fluctuations of the quantum field.
For spacetimes with linear metric perturbations as considered here,
the stochastic correction to the stress tensor has vanishing divergence
(to first order in the metric perturbations).

The stochastic source given by the noise tensor
$\xi_{\mu\nu\alpha\beta}(x)$ (which for this problem has the symmetries
of the Weyl tensor)
is characterized completely by the two point correlation function
which is the noise kernel $N(x-y)$
(the probability distribution for the noise is Gaussian)  
\cite{HuSin,CamVer}
\begin{eqnarray}
      \langle\xi_{\mu\nu\alpha\beta}(x)\rangle_{\xi}
  &=& 0,
      \nonumber \\
      \langle\xi_{\mu\nu\alpha\beta}(x)
      \xi_{\rho\sigma\lambda\theta}(y)\rangle_{\xi}
  &=& T_{\mu\nu\alpha\beta\rho\sigma\lambda\theta}
      \mbox{\rm N}(x-y),
   \label{eq:gaussian correlations}
\end{eqnarray}
where $T_{\mu\nu\alpha\beta\rho\sigma\lambda\theta}$ is
the product of four metric tensors (in such
a combination that the right-hand side of the equation satisfies
the Weyl symmetries of the two stochastic fields on the left-hand side).
Its explicit form is given in \cite{CamVer}
 
If we now take the mean value of equation
(\ref{eq:effective stress tensor}) with respect to the stochastic
source $\xi$ we find that, as a consequence of the noise correlation  
relation,
\begin{equation}
          \langle T^{\mu\nu}_{eff}
          \rangle_\xi
        = \langle T^{\mu\nu} \rangle_q
\end{equation}
and we recover the semiclassical Einstein equations.

The stochastic term $2F^{\mu\nu}$ will produce a stochastic
contribution $h^{st}_{\mu\nu}$ to the spacetime inhomogeneity,
{\it i.e.} $h_{\mu\nu}=h^c_{\mu\nu}+h^{st}_{\mu\nu}$, which we call
metric fluctuations. Considering a flat background spacetime for
simplicity (setting $\a=0$), one obtains (by adopting the harmonic gauge
condition $(h^{st}_{\mu\nu} - {1\over2}\eta_{\mu\nu}h^{st})^{,\nu}=0$)
a linear equation for the metric fluctuations $h^{st}_{\mu\nu}$
\begin{eqnarray}
   \Box h^{st}_{\mu\nu}
        &=& 2\k S^{st}_{\mu\nu},
           \nonumber \\
   S^{\mu\nu}_{st}
        &=& 2F^{\mu\nu}
         =-4\partial_\alpha\partial_\beta\xi^{\mu\alpha\nu\beta},
\end{eqnarray}
The solution of these equations and the computation of the noise correlations
have been given by Campos and Verdaguer in their beautiful paper, from where the
above schematic description is adapted and further details can be found.

We believe this new framework is fruitful for investigation into metric
fluctuations and backreaction effects. We and others \cite{HuSin,CamVer,CCV}
have applied it to study quantum effects in cosmological spacetimes. Work
on black hole spacetimes is just beginning. 
Let us first review what has been done before,  what we regard as
deficient and describe the setup of this problem in our approach.


\section{Metric fluctuations and Backreaction in Semiclassical Gravity}

It is perhaps useful to begin by emphasizing the difference in meaning of
`metric fluctuations' used in our approach and that used by others.  
In the glossary of almost all other authors metric fluctuations have been
used in a test field context,
referring to the two-point function of gravitational perturbations $h_{\m\n}$
in the classical sense or the expectation value of graviton two-point function 
$\langle h_{\m\n}(x) h_{\r\s}(y)\rangle$ in a semiclassical sense --
semiclassical in that the 
background remains classical even though the perturbations are quantized.
It is in a test field context  because one considers gravitational perturbations 
and their two-point functions from a fixed background geometry . This is a useful
concept, but says nothing about backreaction. It is useful as a measure of
the fluctuations in the gravitational field at particular regions of spacetime.
Ford has explored this aspect in great detail. For example, the recent work
of Ford and Svaiter \cite{ForSva} shows that black hole horizon
fluctuations  are much
smaller than Planck dimensions for black holes whose mass exceeds the Planck 
mass. For these black holes they induced that semiclassical derivation 
of the Hawking radiance should remain valid, and that contrary to some recent
claims \cite{CEIMP,Sorkin}, there is no drastic effect near the horizon
arising from metric fluctuations. However, for backreaction considerations, 
where the background spactime metric varies in accordance with the behavior of 
quantum fields present, the graviton 2-point function calculated with respect 
to a fixed background is not the relevant quantity to consider.

In contrast, metric fluctuations $h^{st}_{\m\n}$ (see Eq.(2.11)) in our work 
\cite{CH94,HM3,HuSin} and that of Campos and Verdaguer \cite{CamVer} are
defined for semiclassical gravity in a manifestly backreaction context.
They are classical stochastic quantities arising  from stress tensor
fluctuations in the quantum
fields present and are perhaps important only at  the Planck
scale\footnote{The two point 
function of  gravitons are not stochastic
variables and so in a stricter sense they should not be called  metric
`fluctuations'. To avoid confusion we may at times called our quantities 
$h^{st}_{\m\n}$, induced metric fluctuations.}. We see that they are
derived 
from the noise kernel, which involves 4-point functions  of the gravitons.
It is this
quantity which enters into the fluctuation-dissipation relation -- not the
usual graviton 2 point function -- which encapsulates the  semiclassical
backreaction.  
This is an important conceptual point which has not been duly recognized.

\subsection{Fluctuation-Dissipation Relation Description of Semiclassical 
Backreaction}

With this understanding let us now expound the typical form of
fluctuation-dissipation relation (FDR) in quantum field theory,
starting with the paradigm of quantum Brownian motion (see, for example,
\cite{if}).
Consider a quantum mechanical detector or atom coupled linearly to
a quantized, otherwise free, field which is initially in some quantum
state, pure or mixed 
(typically taken to be a thermal state). After the coupling is switched
on, the atom will ``relax'' to equilibrium over a time scale which depends
on the coupling constant. When the atom reaches equilibrium, the
equilibrium fluctuations of its observables depend  on the quantum
state of the field (for example, if they are thermal fluctuations, the
associated temperature will be the temperature of the field). There are
therefore two relevant processes: dissipation in the atom as it approaches
equilibrium, and fluctuations at equilibrium. These two processes are
generally related by a fluctuation- dissipation relation.

For quantum fields,  let us consider the model of a simplified atom or
detector with internal coordinate $Q$, coupled to a quantized scalar field
$\phi$ via a bilinear interaction with coupling constant $e$: $L_{I}(t) =
e Q(t)\phi(x(t))$, $t$
being the
atom's proper time and $x(t)$ denoting its parametrized trajectory.
It can be shown \cite{HM3,RHK} that the
semiclassical dynamics of $Q$
 is
given by stochastic equations of the form 
\be
\frac{d}{dt}\frac{\partial L}{\partial
\dot{Q}} - \frac{\partial L}{\partial Q} + 2\int^t \gamma(t, s)\dot{Q}(s)
= \xi (t),
\te
where $\xi(t)$ is a stochastic
force arising out of quantum or thermal fluctuations of the field,  and
$L$ is the free Lagrangian
describing the dynamics of the internal coordinate $Q$. Its
correlator is defined as $\langle \xi(t) \xi(t')\rangle = \hbar \nu (t,
t')$. 

The functions $\gamma$ and $\nu$
can be written in terms of two-point functions of the field bath
surrounding the atom. Thus,
\ba
\mu(t,t') &=& \frac{d}{d(t-t')}\gamma(t-t') = \frac{e^2}{2}G(x(t),x(t')) =
-i \frac{e^2}{2} \langle
[\phi(x(t), \phi(x(t'))]\rangle \nn \\ 
\nu(t,t') &=& \frac{e^2}{2}G^{(1)}(x(t),x(t')) =  \frac{e^2}{2} \langle
\{\phi(x(t), \phi(x(t'))\}\rangle , 
\ta   
where $G$ and $G^{(1)}$ are respectively the Schwinger (commutator)  and
Hadamard
(anticommutator) functions of the free field $\phi$ evaluated at two
points on the atom's trajectory. Both functions are evaluated in whatever
quantum state the field is initially in, not necessarily a vacuum state.
Because of the way $\mu$ and $\nu$ enter the equations of motion, they are
referred to as {\it dissipation} and {\it noise} kernels, respectively. It
should be noted that these two kernels, although independent of $Q$,
ultimately determine the rate of energy dissipation of the internal
coordinate $Q$ and its quantum/thermal fluctuations.

The statement of the FDR for such cases is that $\nu$ and $\gamma$ are
related by a linear non-local relation of the form
\be
\label{linresp1}
\nu(t- t') = \int d(s-s')K(t- t', s-s') \gamma(s-s').
\te 
For thermal states of the field, and in $3+1$ dimensions, it can be shown
that 
\be
\label{linresp2}
K(t,s) = \int_{0}^{\infty}\frac{d\omega}{\pi}\omega\coth\left(\ha \beta
\hbar \omega \right) \cos (\omega (t-s)),
\te
where $\beta$ is the inverse temperature. Equations (\ref{linresp1}) and
(\ref{linresp2}) constitute the general form of a FDR. With these ideas  
in mind, 
we will now discuss the two forms of FDR which have appeared in the literature 
in the context of static (this section) and dynamic (next section) black holes.
   
\subsection{Fluctuation and Backreaction in Static Black Holes}

Backreaction in this context usually refers to seeking a consistent  
solution of the 
semiclassical Einstein equation for the geometry of a black hole in  
equilibrium with its
Hawking radiation (enclosed in a box to ensure relative stability).    
Much effort in the
 last 15 years has been devoted to finding a regularized energy-momentum  
tensor for the backreaction calculation. See \cite{JMO} for 
recent status and earlier references.  One important early work on  
backreaction is by 
York \cite{York}, while the most thorough is carried out by Hiscock,  
Anderson et al
\cite{AndHis}.
 
Since the quantum field in such problems is assumed to be in a  
Hartle-Hawking state, 
concepts and techniques from thermal field theory are useful.  Hartle and
Hawking
\cite{HarHaw}, Gibbons and Perry \cite{GibPer} used the periodicity condition of the
Green function on the Euclidean section to give a simple derivation of the Hawking 
temperature for a Schwarzschild black hole.
In the same vein,  Mottola \cite{Mot} showed that in some generalized Hartle-Hawking 
states a FDR exists between the expectation values of the commutator  and
anti-commutator of the energy-momentum tensor 
This FDR is similar to the standard thermal form found in linear response
theory:
\be
\label{fdr}
S_{abcd}(x,
x')=\int_{-\infty}^{\infty}\frac{d\omega}{2\pi}e^{-i\omega(t-t')} \coth \left(\ha 
\beta \omega\right) \tilde{D}_{abcd}({\bf x}, {\bf x'};\omega),
\te
where $S$ and $D$ are the anticommutator and commutator functions of  the
energy-momentum
tensor, respectively, and $\tilde{D}$ is the temporal Fourier transform of $D$.
That is,
\ba
S_{abcd}(x, x') &=& \langle\{\hat{T}_{ab}(x),\,\hat{T}_{cd}(x')\}\rangle_{\beta} 
\nn \\
D_{abcd}(x, x') &=&
\langle[\hat{T}_{ab}(x),\,\hat{T}_{cd}(x')]\rangle_{\beta}.
\ta 
He also identifies the two-point function $D$ as a dissipation
kernel by relating it to the 
time rate of change of the energy density when the metric is slightly perturbed. 
Thus, Eq.(\ref{fdr}) represents a {\it bona fide} FDR relating the  
fluctuations of a certain quantity 
(say, energy density) to the time rate of change of the very same quantity.

However, this type of FDR has rather restricted significance as it is
based on the assumption of a fixed background spacetime (static in this
case) and state (thermal) of the matter field(s).
It is not adequate for the description of backreaction where the spacetime
and the state of matter are determined in a self-consistent manner by
their dynamics and mutual influence.  One should  look
for a FDR for a parametric family of metrics (belonging to a general class)
and a more general state of the quantum matter (in particular, for
the Unruh state). We expect the derivation of such an
FDR will be far more complicated than the simple case where the Green
functions are periodic in imaginary time, and where one can simply take
the results of linear response theory almost verbatim.

Even in this simple case, it is noteworthy that there is a small
departure from standard linear response theory for quantum systems. This
arises from the observation that the dissipation kernel 
in usual linear response analyses is given by a two-point
commutator function of the underlying quantum field, which is independent
of the quantum state for free field theory. In this case, we are still
restricted to free fields in a curved background. However, since the
dissipation now depends on a two-point function of the stress-tensor,
it is a four-point function of the field, with appropriate derivatives and
coincidence limits. This function is, in general, state-dependent.
We have seen examples from related cosmological backreaction problems
where it is possible to explicitly relate the dissipation to particle creation
in the field, which is definitely a state-dependent process. For the
black-hole case, this would imply a quantum-state-dependent damping of
semiclassical perturbations.  The temperature dependence of the  
anticommutator function 
$S$ as displayed in the fluctuation-dissipation relation of Eq. (\ref{fdr}) is 
therefore misleading, because the 
commutator function $D$ can also be temperature-dependent. 

To obtain a
causal FDR for states more 
general than the Hartle-Hawking state, one needs to use the in-in 
(or Schwinger-Keldysh) formalism applied to a class of quasistatic metrics
(generalization of York \cite{York}) and proceed in a way leading to
the noise kernel similar to that illustrated in Sec. 2. The calculation  
of 4-point functions 
of a thermal scalar field in black hole spacetimes \cite{HPR} and  the
derivation of such an FDR  are in progress \cite{CHR}.

\section{Fluctuations and Backreaction in Dynamical Black Holes}

\subsection{Quasistationary Approximation and Bardeen Metric}

Backreaction for dynamical (collapsing) black holes are much more
difficult 
to treat than static ones, and there are fewer viable attempts. 
For situations with black hole masses much greater than the Planck mass, 
one important work which captures the overall features of dynamical
backreaction is that by Bardeen \cite{Bardeen}, who, using a
generalization of a classical
model geometry (Vaidya metric for stars with outgoing perfect fluid),
argued that 
the mass of a radiating black hole decreases at a rate given
by its luminosity, as expected from energy considerations. That is,
\be
\label{lum}
\frac{dM}{dt}= - L.
\te
In particular, for a black hole emitting Hawking radiation,
the luminosity goes as $L =\a \hbar M^{-2}$, $M$ being the black hole
mass, and $\a$ some constant.
Far from the horizon ($r > {\cal O}(6M)$), Bardeen's geometry takes
the form
\be
ds^2 = - \left(1-\frac{2m(u)}{r}\right)du^2 - 2du\,dr +r^2d\O ^2,
\te
with
\be
m(u) = \int^{u}du\, L(u),
\te
$L(u)$ being the Hawking luminosity, and $m(u)$ the Bondi mass.
With this Bardeen argued that the semiclassical picture of black hole
evaporation should hold until the black hole reaches Planck size.

More recently Parentani and Piran \cite{ParPir},
using a spherically symmetric geometry and a simplified scalar field model
which neglects the potential barrier, carried out a numerical integration
of the coupled quantized scalar field and semiclassical Einstein equations,
and showed that the solution of the semiclassical theory in this model is
the geometry described by Bardeen. Using the same model, Massar \cite{Mas}
recently showed that the emitted particles continue to have a
thermal distribution with a time-dependent Hawking temperature
$(8\pi M(t))^{-1}$. We refer readers to the latter work for details.

With this as a backdrop, the goal of our current program is to \cite{HRS}\\
1) derive a fluctuation-dissipation relation embodying the backreaction
for this quasistationary regime;\\
2) apply the stochastic field-theoretic formalism to the near-Planckian
regime and derive an Einstein-Langevin equation for the dynamical metric
including the effects of induced metric fluctuations.\\

\subsection{Fluctuation-Dissipation Relation of Candelas and Sciama}

On the first issue, historically Candelas and Sciama \cite{CanSci}
first proposed a fluctuation-dissipation relation 
for the depiction of dynamic black hole evolutions.
But as we will point out, their relation does not include backreaction in
full and is not a FDR in the correct statistical mechanical sense.

As a starting point they considered the {\it classical} relation, 
due to Hartle and Hawking \cite{HarHaw73},  between energy flux
transmitted across the horizon of a perturbed black hole and the shear
\footnote{The analysis of Candelas and Sciama holds for a Kerr black hole. 
However, we  simplify this to the Schwarzschild case in the interest of  
clarity; 
no qualitative features are lost in this simplification.}:
\be
\frac{d^2E}{dt d\Omega} =  \frac{M^2}{\pi} \mid \sigma (2M) \mid^2,
\te
where $\sigma (2M)$ is the perturbed shear of the null congruence which
generates the future horizon.

In turn, the dissipation of horizon area with respect to the advanced null
coordinate is related to the energy flux across the horizon, and the
above equation becomes (see, for example \cite{HawLH})
\be
\label{csfdr}
\frac{dA}{dv} = 4M \int_{H} \mid \sigma \mid^2 dA,
\te
the integral being performed over the horizon.

The  classical formula above immediately suggests a
fluctuation-dissipation description: the dissipation in area is related
linearly to the squared absolute value of the shear amplitude.
This description is even more relevant when the gravitational
perturbations are quantized. Then the integrand of the right-hand-side of
Eq.(\ref{csfdr}) is $\langle \sigma^{\ast} \sigma \rangle$, the
expectation value being taken with respect to an appropriate quantum
state. Candelas and Sciama choose this state to be the Unruh vacuum,
arguing that it is the vacuum which approximates best a flux of
radiation from the hole at large radii.

The details of this expectation value are given in \cite{CanSci}. Here we
simply note  that, with the substitution of this quantity in
(\ref{csfdr}), the left-hand-side of that equation now represents the
dissipation in area due to the Hawking flux of gravitational radiation,
and the right-hand-side comes from pure quantum fluctuations of gravitons 
(as opposed to semiclassical fluctuations of gravitational perturbations, 
which are induced by the presence of quantum matter). It is tempting to
regard this as a quantum FDR characteristic of the Hawking process, as do 
the authors of \cite{CanSci}.

However, Eq.(\ref{csfdr}) is not a FDR in a truly statistical
mechanical sense because it does not relate
dissipation of a certain quantity (in this case, horizon area) to the
fluctuations of {\it the same quantity}. 
To do so would require one to compute the two point function of the
area, which, in turn, is a four-point function of the graviton field, and
intimately related to a two-point function of the stress tensor. The
stress tensor is the true ``generalized force'' acting on the spacetime
via the equations of motion, and the dissipation in the metric must
eventually be related to the fluctuations of this generalized force for
the relation to qualify as an FDR.  The calculation of 4-point functions
of the  metric perturbations $h_{ab}$ and the correct FDR is currently  
under investigation \cite{HPR,HRS}

\subsection{Bekenstein's theory of black hole fluctuations}

The importance attributed to the correlation of mass function
was a central point in Bekenstein's theory of black hole fluctuations
\cite{Bek}.
Because it bears some similarity in conceptual emphasis to our approach
we'd like to refresh the reader's memory of this work. We will also
comment on the basic difference from our approach.

Bekenstein  considered the mass fluctuations  (and fluctuations
of other parameters) of an {\it isolated} black hole due to the
fluctuations in the radiation emitted by the hole, and considers the
question of when such fluctuations can be large.  For simplicity, only
mass fluctuations are considered. The basic assumption is that the black
hole mass $M(t)$ is a stochastic function with some probability
distribution due to the stochastic emission of field quanta, and
furthermore, that energy is conserved during the stochastic emission of
quanta. As we shall see, this latter assumption leads to some startling
predictions. With these assumptions, one may express the black hole mass at
some time $t + dt$ in terms of the mass at an earlier time $t$ by the
equation
\be
\label{stoc}
M(t+dt) = M(t) - m(dt),
\te
where $m(dt)$ is the energy taken away by radiation in time $dt$.
Averaging the above equation yields
\ba
\label{stoc2}
\frac{d}{dt}\langle M\rangle &=& -\frac{m(dt)}{dt} = -\langle L \rangle
\nn \\
&=& -\a \hbar \langle M^{-2} \rangle,
\ta
where Eq. (\ref{lum}) with the Hawking luminosity is used here.
Furthermore, squaring  Eq. (\ref{stoc}) before taking the average yields,
\be
\label{stoc3}
\frac{d}{dt}\langle M^2 \rangle = 2\a \hbar \langle M^{-1} \rangle + \b
\hbar ^2 \langle M^{-3} \rangle,
\te
the second term in the above expression being obtained by Bekenstein from
an approximate expression for the energy distribution of quanta
emitted in time $dt$. 

Defining $\sigma_M = \langle M^2 \rangle - \langle M \rangle^2 $, and
using the moment expansions
\ba
\langle M^{-1} \rangle &=& \frac{\langle M^2 \rangle}{\langle M \rangle^3}
+ {\cal O}(\langle M^{3} \rangle) \nn \\
\langle M^{-2} \rangle &=& -\frac{2}{\langle M \rangle^2} + \frac{3\langle
M^2 \rangle}{\langle M \rangle^4} + {\cal O}(\langle M^{3} \rangle) \nn \\
\langle M^{-3} \rangle &=& -\frac{5}{\langle M \rangle^3} + \frac{6\langle
M^2 \rangle}{\langle M \rangle^5} + {\cal O}(\langle M^{3} \rangle), 
\ta
Eqs. (\ref{stoc3}) and (\ref{stoc2}) together imply
\be
\frac{d}{dt} \sigma_M = \frac{\hbar}{\langle M \rangle^3}\left(-4\a
\sigma_M + \b \hbar (1+6\sigma_M)\right).
\te
The above differential equation for $\sigma_M$ possesses the approximate
solution
\be
\sigma_M \sim \c\, \hbar \left(\frac{M_0^4}{\langle M \rangle^4} -1\right),
\te
$\c$ being a constant related to $\a $ and $\b $, and $M_0^4$ an
integration constant, to be interpreted as the mass when it is sharply
resolved (i.e. when $\sigma_M =0$). As pointed out by Bekenstein, one of 
the many consequences of this solution is that the fluctuations $\sigma_M$
can grow as large as $\langle M\rangle^2$ for $\langle M \rangle =
M_c \sim \hbar^{1/6} M_0^{2/3}$. According to this picture, therefore,
depending on the initial mass,
mass fluctuations can grow large far before the Planck scale. Once the
critical mass $M_c$ is reached, the dynamics of the black hole differs
drastically from the standard semiclassical picture, because the equation
for
$\langle M \rangle$, Eq. (\ref{stoc2}), is itself driven by the
fluctuations in mass. 

A crucial assumption for the validity of such a
scenario for black hole evaporation is, of course, the stochastic energy
balance equation (\ref{stoc}) and the related Eq. (\ref{stoc2}). It is
possible that different assumptions for the stochastic dynamics lead to
drastically different conclusions about the late stages of the evaporation
process. The prediction of Bekenstein's theory is at variance with Bardeen's
in which the semiclassical picture of black hole evaporation remains valid
until the hole reaches Planck size.
Our stochastic semiclassical gravity theory would also support Bardeen's  
scenario
as the fundamental stochastic (Einstein-Langevin) equation for the black hole
mass is expected to take the form
\be
\label{EL}
\frac{dM}{dt} = -\frac{\a \hbar}{\langle M \rangle^2} + \xi (t),
\te
where $\xi(t)$ is a stochastic term with vanishing average value. If such
an equation were to hold, the average mass would be independent of the
fluctuations, while the fluctuations would be slaved to the dynamics of
the average mass, where backreaction will be incorporated in a  
self-consistent
manner. As we showed in Sec. 2 this type of behavior of the mean field
(semiclassical metric) and fluctuations is also found to occur in the
treatment of backreaction in cosmological spacetimes.

We anticipate an Einstein-Langevin equation of the form (\ref{EL}) for
the description of black hole evaporation in contrast to Bekenstein's
equations of the form (\ref{stoc2}). 
The Einstein-Langevin equation when averaged
over the probability distribution for the noise $\xi$ should yield 
a semiclassical Einstein equation of the Bardeen type as a mean field theory.

\section{Prospects}

Here we have discussed some representative work related to ours  on
semiclassical
black hole fluctuations and backreaction (other noteworthy proposals
include that of 'tHooft \cite{tHooft}, and work on 2D dilaton gravity  
\cite{BPP}).
We have also sketched our approach, and marked out some important
points of departure. This include 1) metric fluctuations and their
role in backreaction -- our definition of (induced) metric fluctuations
is in terms of graviton 4 point functions; 2) the true statistical  
mechanical meaning of
the fluctuation-dissipation relation and its embodiment of the  
backreaction effects.
Our formulation testifies to the existence of a stochastic regime in
semiclassical gravity where the dynamics of spacetime is governed by an
Einstein-Langevin Equation which
incoporates metric fluctuations induced by quantum field processes.
We wish to explore the physics in this new stochastic regime, including
possible phase transition characteristics near the Planck scale and
its connection with low energy string theory predictions.

For the implementation of this program currently we are engaged in\\
a) setting up the CTP effective action for the quasistatic and dynamic
 cases \cite{CHR,HRS},\\
b) computing the fluctuations of the energy momentum tensor for 
fields both in the Hartle-Hawking state and the Unruh state \cite{HPR},  
and\\
c) exploring the interior solution of a collapsing black hole  \cite{HLPS}
as this bears 
closer resemblance to a cosmological problem (Kantowski Sachs universe) 
\cite{AndHis}.

This program will take a few years to fruition and we hope to report on some 
results in Vishu's 65th birthday celebration.\\

{\bf Acknowledgement} This work is supported in part by NSF grant PHY94-
21849 to the University of Maryland and PHY95-07740 to the University of
Wisconsin-Milwaukee. Part of this work was reported by BLH at the Second
Symposium on Quantum Gravity in the Southern Cone held in January 1998 
in Bariloche, Argentina.


\end{document}